ARTICLE   OPEN

# Machine-learned epidemiology: real-time detection of foodborne illness at scale

Adam Sadilek[1], Stephanie Caty[2], Lauren DiPrete[3], Raed Mansour[4], Tom Schenk Jr[5], Mark Bergtholdt[3], Ashish Jha[2,6], Prem Ramaswami[1] and Evgeniy Gabrilovich[1]

Machine learning has become an increasingly powerful tool for solving complex problems, and its application in public health has been underutilized. The objective of this study is to test the efficacy of a machine-learned model of foodborne illness detection in a real-world setting. To this end, we built FINDER, a machine-learned model for real-time detection of foodborne illness using anonymous and aggregated web search and location data. We computed the fraction of people who visited a particular restaurant and later searched for terms indicative of food poisoning to identify potentially unsafe restaurants. We used this information to focus restaurant inspections in two cities and demonstrated that FINDER improves the accuracy of health inspections; restaurants identified by FINDER are 3.1 times as likely to be deemed unsafe during the inspection as restaurants identified by existing methods. Additionally, FINDER enables us to ascertain previously intractable epidemiological information, for example, in 38% of cases the restaurant potentially causing food poisoning was not the last one visited, which may explain the lower precision of complaint-based inspections. We found that FINDER is able to reliably identify restaurants that have an active lapse in food safety, allowing for implementation of corrective actions that would prevent the potential spread of foodborne illness.

npj Digital Medicine (2018)1:36 ; doi:10.1038/s41746-018-0045-1

## INTRODUCTION

In the 1800s, John Snow had to go door to door during an epidemic of cholera to uncover its mechanisms of spread.[1] He recorded where people were getting their drinking water from in order to pinpoint the source of the outbreak. Here we scale up this approach using machine learning to detect potential sources of foodborne illness in real time. Machine learning has become an increasingly common artificial intelligence tool and can be particularly useful when applied to the growing field of syndromic surveillance. Frequently, syndromic surveillance depends upon patients actively reporting symptoms that may signal the presence of a specific disease.[2,3] In recent years, syndromic surveillance has also begun to include passively collected information, such as information from social media, which can also lend insight into potential disease outbreaks.[4–6] In this study, we use such observational data to identify instances of foodborne illness at scale.

Frequently, in the United States and elsewhere, efforts to combat disease outbreaks still rely on gathering data from clinicians or laboratories and feeding this information back to a central repository, where abnormal upticks in prevalence can be detected.[7,8] For instance, when foodborne illnesses occur in the United States, determining an outbreak is dependent upon either complaints from large numbers of patients or receipt of serological data from laboratory tests.[9] These processes can be slow and cumbersome and often lead to a delayed response, allowing for further spread of disease.[10] Having the ability to track and respond to outbreaks in real time would be immensely useful and potentially lifesaving.

Here we sought to test the efficacy of a machine-learned model that uses aggregated and anonymized Google search and location data to detect potential sources of foodborne illness in real time. Our primary goal was to use this model to identify restaurants with potentially unsafe health code violations that could contribute to foodborne illness spread, with the hypothesis that our model would be able to more accurately identify a restaurant with serious health code violations than systems currently in place. We find that such an approach can lead to a greater than threefold improvement in identifying potentially problematic venues over current approaches, including a 68% improvement over an advanced complaint-based system that already utilizes Twitter data mining. Our results provide evidence that this type of tool can be used by health departments today to more rapidly pinpoint and investigate locations where outbreaks may be occurring. This model can be expanded by public health departments to reduce the burden of foodborne illness across the United States, and can also be expanded to assist in monitoring a variety of other diseases globally.

### FINDER machine-learning methodology

Here we introduce a machine-learned model called FINDER (Foodborne IllNess DEtector in Real time), which detects restaurants with elevated risk of foodborne illness in real time. The model leverages anonymous aggregated web search and location data and ensures that specific findings cannot be attributed to individual users. We call this approach machine-learned epidemiology. It complements existing approaches to identifying illnesses with new real-time signals available at large scale.

[1]Google Inc., 1600 Amphitheatre Parkway, Mountain View, CA 94043, USA; [2]Harvard T.H. Chan School of Public Health, 42 Church St, Cambridge, MA 02135, USA; [3]Southern Nevada Health District, 280 S Decatur Blvd, Las Vegas, NV 89107, USA; [4]Chicago Department of Public Health, 333 S State St #200, Chicago, IL 60604, USA; [5]Chicago Department of Innovation and Technology, 333 S State St #420, Chicago, IL 60614, USA and [6]Veterans Affairs Boston Healthcare System, 150 S Huntington Ave, Boston, MA 02130, USA
Correspondence: Ashish Jha (ajha@hsph.harvard.edu)
These authors contributed equally: Adam Sadilek, Stephanie Caty.

FINDER applies machine learning to Google search and location logs to infer which restaurants have major food safety violations, which may be causing foodborne illness. This anonymous and aggregated logs data comes from users who opted to share their location data, which already enables other applications, such as estimates of live traffic.

Our method first identifies queries indicative of foodborne illness, and then looks up restaurants visited in aggregate by the users who issued those queries, leveraging their anonymized location history. FINDER then calculates, for each applicable restaurant, the proportion of users who visited it and later showed evidence of foodborne illness in their searches. Notably, in most previous work, a user's location is only known if she searched or posted a message from the location.[11,12] In contrast, our data source is much more comprehensive, allowing us to reliably infer previously visited locations, regardless of whether the user took any action there.

The key challenge is the inherent noise and ambiguity of individual search queries. For example, the query [diarhea] could be related to food poisoning, but also contains a typo and does not convey information about the details of the symptom (e.g., what type of diarrhea, is it experienced by the user or her family member). We solve this challenge with a privacy-preserving supervised machine-learned classifier, which leverages a collection of signals beyond the query string itself, such as search results shown in response to the query,[13] aggregated clicks on those results, and the content of the opened web pages. The resulting classifier has high accuracy in identifying queries related to food poisoning, achieving area under the ROC curve of 0.85, and F1 score of 0.74 in evaluation with three independent medical doctors and separately with three non-medical professionals rating each query. Note that an individual affected by foodborne illness starts feeling symptoms with certain delay (incubation period) after the infection has occurred. While FINDER processes log data in real time, confident inference can only be drawn after incubation period has elapsed for an initial cohort of affected patrons.

Application of FINDER in two cities

In order to test the efficacy of FINDER, we deployed the model in Las Vegas, Nevada and Chicago, Illinois. Every morning, each city was provided with a list of restaurants in their jurisdiction that were automatically identified by FINDER. The health department in each city would then dispatch inspectors (who were unaware of whether the inspection was prompted by FINDER or not) to conduct inspections at those restaurants to determine if there were health code violations. In addition to FINDER-initiated inspections, the health departments continued with their usual inspection protocols. The results of the latter inspections were used as a comparison set, with three comparison groups: all inspected restaurants not prompted by FINDER (referred to as BASELINE below), and two subsets thereof—complaint-based inspections (COMPLAINT) and routine inspections (ROUTINE).

We labeled the restaurants as safe or unsafe based on the outcome of the inspection results and report the accuracy of identifying an unsafe venue across the various comparison groups (FINDER, BASELINE, COMPLAINT, and ROUTINE). Restaurants that received a grade reflective of any sort of serious health code violation were designated unsafe. For a complete description of safe/unsafe criteria, see Supplementary Text. We also broke the results down by the risk level of each venue. This study was designated as non-human subjects research by the Harvard T.H. Chan School of Public Health Institutional Review Board.

Table 1. Number of inspections conducted during the experimental time period

|  | FINDER | BASELINE |
|---|---|---|
| Total | 132 | 10,786 |
| Las Vegas | 61 | 4977 |
| Chicago | 71 | 5809 |
| Complaint-driven | N/A | 1291 |
| Routine | N/A | 4518 |
| Risk level[a] |  |  |
| High (% of total) | 84 (63.6%) | 5702 (52.9%) |
| Medium (%) | 39 (29.6%) | 2325 (21.6%) |
| Low (%) | 9 (6.8%) | 2759 (25.6%) |

[a]$p$ value for difference in risk distribution between FINDER and BASELINE <0.001, from $X^2$-test

FINDER was deployed in Las Vegas between May and August 2016; during that period a total of 5038 inspections were completed, 61 of which were prompted by FINDER (Table 1). A similar deployment occurred in Chicago between November 2016 and March 2017, where 5880 inspections were completed, 71 of which were prompted by FINDER. Of the inspections not attributed to FINDER, 1291 inspections were driven by complaints through the existing systems in Chicago (Table 1).

RESULTS

Detection of unsafe restaurants

We assessed the accuracy of FINDER's predictions by comparing the fraction of unsafe restaurants it identified to the fraction of unsafe venues found in all the other restaurant inspections conducted during the experimentation period (BASELINE), as well as the fraction of unsafe venues found in the two subgroups, COMPLAINT and ROUTINE.

Of all the restaurants identified by FINDER, 52.3% were deemed unsafe upon inspection, compared to 24.7% for BASELINE restaurants (Table 2). We used binomial logistic regression to determine the odds ratio of being unsafe for restaurants in the FINDER and BASELINE groups. The former were 3.06 times (95% CI: 2.14–4.35) as likely to be unsafe as the latter, when accounting for restaurant risk level and city in our models ($p < 0.001$, Table 2). When stratified by restaurant risk level, FINDER restaurants were more likely to be designated unsafe across all risk designations, however the odds of being identified by FINDER as unsafe was higher in lower risk-level restaurants than in high risk-level restaurants (Table 2). Importantly, this suggests that a priori determination of the restaurant risk level might not necessarily reflect the true level of risk at the venue.

Comparison to complaint-based inspections

We did not examine complaint-based inspections from Las Vegas for two reasons. First, in that city complaints are handled differently from routine inspections in that complaints trigger a very focused investigation based on the nature of the complaint (as opposed to a comprehensive evaluation of food safety at the establishment, as in Chicago). Second, the transient nature of Las Vegas restaurant patrons, many of whom are visitors from elsewhere, means that the number of complaints received is very low, with only 15 complaints being reported during the experimental time period.

Therefore, we focused only on complaints from Chicago. We found that the overall rate of unsafe restaurants among those detected by FINDER in Chicago was 52.1%, whereas the rate of unsafe restaurants in COMPLAINT inspections was 39.4% (Table 2). Adjusting for venue risk level, we found that across all restaurants, the odds ratio that a FINDER restaurant is unsafe was 1.68 times (95% CI: 1.04–2.71) as high as COMPLAINT inspections ($p = 0.03$, Table 2). Across all restaurant risk levels, FINDER restaurants were more likely to be given an unsafe designation than COMPLAINT restaurants (Table 2).

Table 2. Ability of FINDER to detect unsafe restaurants as compared to BASELINE rate and with subcategories of the baseline inspections, including complaint-based inspections that occurred in Chicago and routine inspections from both Chicago and Las Vegas

|  | FINDER $n = 132$ | BASELINE $n = 10{,}786$ | Odds ratio[a] [95% CI] | p-value |
|---|---|---|---|---|
| Overall, number unsafe (%) | 69 (52.3%) | 2662 (24.7%) | 3.06 [2.14–4.35] | <0.001 |
| Risk level |  |  |  |  |
| High, number unsafe (%) | 42 (50.0%) | 1909 (33.5%) | 1.98 [1.28–3.05] | 0.002 |
| Medium, number unsafe (%) | 23 (59.0%) | 536 (23.1%) | 5.50 [2.83–10.72] | <0.001 |
| Low, number unsafe (%) | 4 (44.4%) | 217 (7.9%) | 7.35 [1.79–30.13] | 0.006 |
| Comparison of FINDER to complaint-based inspections |  |  |  |  |
|  | FINDER $n = 71$ | COMPLAINT $n = 1291$ |  |  |
| Overall, number unsafe (%) | 37 (52.1%) | 508 (39.4%) | 1.68 [1.04–2.71] | 0.03 |
| Risk level |  |  |  |  |
| High, number unsafe (%) | 27 (47.4%) | 374 (39.4%) | 1.38 [0.81–2.36] | 0.24 |
| Medium, number unsafe (%) | 9 (75.0%) | 115 (39.3%) | 4.64 [1.23–17.51] | 0.02 |
| Low, number unsafe (%) | 1 (50.0%) | 19 (38.8%) | 1.58 [0.09–26.78] | 0.75 |
| Comparison of FINDER to routine inspections |  |  |  |  |
|  | FINDER $n = 132$ | ROUTINE $n = 9495$ |  |  |
| Overall, number unsafe (%) | 69 (52.3%) | 2,154 (22.7%) | 3.16 [2.22–4.51] | <0.001 |
| Risk level |  |  |  |  |
| High, number unsafe (%) | 42 (50.0%) | 1531 (32.2%) | 2.07 [1.35–3.20] | 0.001 |
| Medium, number unsafe (%) | 23 (59.0%) | 424 (20.9%) | 5.52 [2.84–10.76] | <0.001 |
| Low, number unsafe (%) | 4 (44.4%) | 199 (7.3%) | 7.65 [1.90–30.89] | 0.004 |

[a]Odds ratios from binomial logistic regressions

Comparison to routine inspections
Finally, we compared the precision of FINDER to that of ROUTINE inspections (in both cities), where a venue gets inspected every 6–24 months depending on jurisdiction. The overall rate of unsafe restaurants detected by FINDER was 52.3%, whereas the overall rate of detection of unsafe restaurants in routine inspections was 22.7% (Table 2). Using a binomial logistic regression adjusting for city and risk level, we found FINDER restaurants to be 3.16 times as likely to be unsafe as ROUTINE restaurants (95% CI: 2.22–4.51). FINDER restaurants were more likely to be designated unsafe than ROUTINE restaurants across all risk-level classifications (Table 2).

FINDER has several advantages over the existing inspection mechanisms. Compared to routine inspections, FINDER has a much higher precision rate of identifying unsafe restaurants, and it can discover health violations that might not be found by traditional protocols. Compared to complaint-based inspections, FINDER still has a greater ability to identify restaurants with significant health code violations and is universal, whereas complaints are fairly scarce (in Chicago, only 22% of inspections were based on complaints). Additionally, as we show below, complaints are often misguided as they attribute illness to the wrong venue.

Table 3. Violation counts

|  | FINDER[a] $n = 132$ | BASELINE[a] $n = 5848$ | p-value |
|---|---|---|---|
| Critical violations | 0.40 | 0.21 | 0.001 |
| Major violations | 0.74 | 0.56 | 0.04 |

[a]Adjusted mean violation count, accounting for city and risk level, calculated using linear regressions

Detection of violations
We next examined whether restaurants identified by FINDER are likely to have more serious safety violations compared to those in the BASELINE group. In both cities, we obtained the number of violations identified in each restaurant inspection. We examined the severity of the violations, and focused on the critical and major violations (see Supplementary Tables S1 and S2 for full list of violations). We used a linear regression model that adjusted for city and restaurant risk level to calculate an adjusted mean number of critical and major violations for FINDER and for BASELINE restaurants. FINDER-identified restaurants had a greater number of critical violations (0.40 vs 0.21, $p = 0.001$, Table 3) and major violations (0.74 vs 0.56, $p = 0.04$, Table 3) than BASELINE restaurants.

Location attribution
Finally, we found that FINDER can better attribute the location of foodborne illness to a specific venue than individual reports from customers generally do. For restaurants identified by FINDER, we focused on the customers who later searched for terms indicative of foodborne illness, and then analyzed their entire sequence of prior restaurant visits.

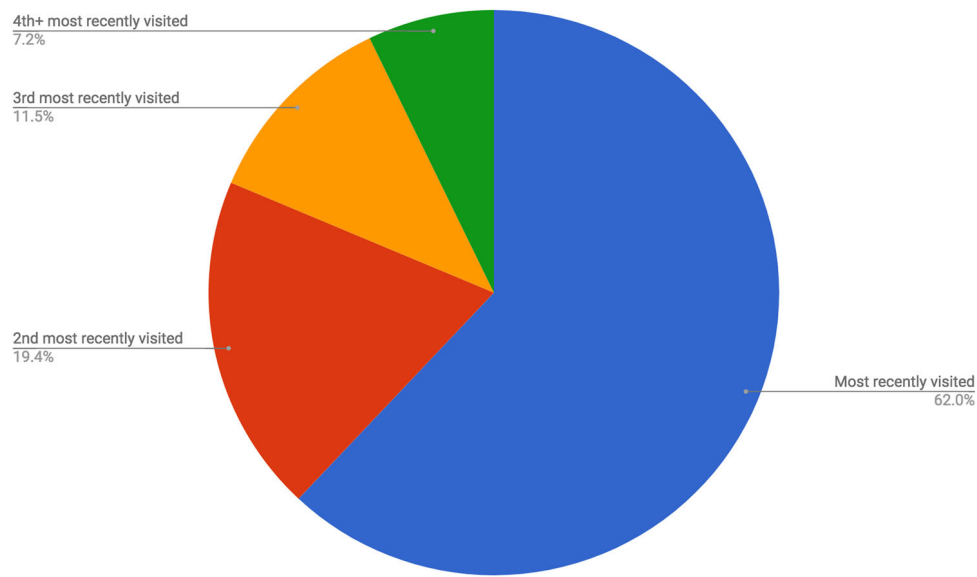

**Fig. 1** Frequency with which illness can be attributed to recently visited restaurants, among FINDER restaurants. $N = 132$

Among all restaurant visits that FINDER associated with a foodborne illness, users appear to have contracted the illness at the restaurant they visited most recently 62% of the time (Fig. 1). However, in the other 38% of cases, their illness was likely caused by another restaurant, which they had visited earlier, given the relative signal strength of the latter restaurant. Specifically, for 19.4% of users, this was the penultimate restaurant visited, for 11.5% it was the third to last restaurant visited, and for 7.2% it was the fourth to last or even an earlier visited restaurant. Previous research shows that people tend to blame the last restaurant visited, and therefore may be likely to file a complaint for the wrong restaurant.[14,15] The FINDER approach is more robust than individual customer complaints, as it aggregates information from numerous people who visited the venue.

## DISCUSSION

After deploying FINDER in two major US cities, we found that FINDER can more precisely identify restaurants with significant health code issues than traditional methods for selecting restaurants for inspection, and with more precision than even complaint-based inspections. Importantly, we found that complaint-based inspections may often be mistargeted. Our findings suggest that large-scale real-time monitoring systems offer a promising way to bolster food inspection efforts and reduce foodborne illness in a large population.

FINDER addresses many gaps that currently exist in this type of syndromic surveillance. First, many syndromic surveillance technologies do not have the capability to geographically pinpoint the specific location or venue where signal is originating from. Even complex systems, such as HealthMap, focus on broader geographic districts.[16] FINDER, on the other hand, is able to use not only real-time geographic location, but also can access recent historical locations to better localize the signal to the most likely epicenter (in FINDER's case, a specific restaurant), rather than to a location where the infection was recorded (e.g., hospital address where patient was treated). Furthermore, thanks to its use of anonymized logs data, FINDER is not subject to patrons' recall bias, which is present in most other systems.

Additionally, many syndromic data surveillance signals are difficult to validate.[3] The accuracy of most disease prediction modeling using online data is evaluated using aggregated past observations,[11,17,18] with the notable exception of nEmesis, a system that used geo-tagged public Twitter messages to detect foodborne illness clusters on a small scale.[12] In contrast, we were able to validate our model through actual health inspections following a standard professional protocol. Notably, we found that our model can more precisely identify restaurants with food safety violations than the system in Chicago, which has one of the most advanced monitoring programs in the nation as it employs social media mining and illness prediction technologies to target their inspections.[19]

Web search queries and online big data have been used before in public health research, most notably in Google Flu Trends.[20–25] The latter model tracked the proportion of 45 manually selected queries over all queries from a given region. These queries were not machine learned and therefore the model was potentially more susceptible to drift and noise over time.[26] In contrast to Google Flu Trends, FINDER uses machine learning to identify the infinite variety of ways in which symptoms of foodborne illness can be described in natural language. Our Web Search Model (WSM, explained in Methods) further improves the understanding of individual queries using search results returned for them. Moreover, Google Flu Trends estimated query volume rather than user volume as we do in this work. These factors together allow us to more reliably estimate incidence rates in a robust and accurate way.

Our study is not without limitations. Specifically, we used data from Google search users, which is a subset of the entire population. However, there is nothing unique to Google in our approach, and other search engines that have location history can create similar algorithms and likely achieve comparable results. Second, although FINDER has a high positive predictive value, it did not detect all the venues with violations flagged through the traditional complaint-driven channels. This is due, in part, to the relatively small number of FINDER restaurants inspected, owing to the limited bandwidth provided to us by city/county health departments, which restricted the number of inspections FINDER could suggest in a given city. To this end, we applied an arbitrary cutoff of signal strength to identify problematic restaurants to send to county health officials, which resulted in a small sample size given time and resource constraints. However, we are able to rank the relative risk of all restaurants in a city, and thus can provide more substantial lists of problematic restaurants to cities in the future to further aid in prioritization of inspections.

Furthermore, while the model will continue to improve to better detect restaurants with health code violations, there is clear evidence that FINDER is best used as a supplement to other methods that cities use but is not yet ready to replace the broader inspection scheme. We further observe that by law every restaurant has to be inspected once or twice a year (depending on the jurisdiction), and FINDER augments traditional inspection mechanisms by suggesting good times to perform the inspections (when the risk of foodborne illness at a venue is high).

Admittedly, the implementation of FINDER does incur some costs for city or county health departments that implement it, in terms of personnel time spent working with and responding to the signal. However, these costs should be considered in the context of cities already having inspectors visiting restaurants, most of which do not have a problem (i.e., higher false positive rate) and if FINDER is able to help cities prioritize inspections, it can be more efficient. It should also be noted that while FINDER does increase the overall accuracy with which county health departments are able to identify restaurants with serious health code violations, there were times when FINDER's predictions were not accurate, and thus inspectors were sent to inspect the restaurant but did not find serious safety issues. This may raise concerns about allocating inspection resources when FINDER predictions are incorrect. Many times, the FINDER-prompted inspection was the first health department inspection of the year for the restaurant. This means that even if it were a misclassification, the visit itself allowed the health department to meet the legal mandate of an inspection for the year. By law, every restaurant has to be inspected once or twice a year (depending on the jurisdiction). Thus, FINDER's moving up the inspection timeline did not require the use of any additional resources. Redundancy or waste issues only arise when FINDER misclassifies a restaurant that has already been inspected. In these cases, individual health departments may choose to shift the priority of restaurants that have already met their inspection requirement.

Overall, the costs of deploying FINDER should be weighed against the costs of foodborne illnesses that would be or continue to be missed. Anecdotally, both Chicago and Las Vegas reported that incorporating FINDER into their current systems required some upfront time and resources, but that soon thereafter, it did not require much additional effort to maintain, and provided valuable insight into inspection priorities.

In conclusion, we found that FINDER can be integrated into existing inspection protocols quickly and at very low financial costs. If deployed broadly, FINDER can potentially be an important part of a national effort to reduce the burden of foodborne illness. Once the model is widely deployed, the feedback from actual inspections can be used as additional training data to further improve the model.

## METHODS

### Experimental design

FINDER is a machine-learned model for real-time foodborne illness detection. To determine the ability of FINDER to detect potentially unsafe restaurants, we introduced FINDER into two local health departments in Chicago and Las Vegas. In each city, FINDER-identified restaurants were inspected following the same protocol used in other restaurant inspections. The results of the FINDER inspections were then compared to the overall baseline inspection results, as well as to two subsets of baseline inspections, complaint-based inspections, and routine inspections that are conducted at certain time intervals.

Analyses were further stratified by restaurant risk level. Both Chicago and Las Vegas designate risk levels for all food establishments, based on the type of establishment and level of food preparation. In each city, these risk categories included low risk (restaurant only handles and serves ready-to-eat ingredients), medium risk (restaurant cooks raw food for same-day service), or high risk (restaurant cooks, cools, and then reheats food on a later date). Of all FINDER-identified restaurants across both cities, 84 (63.6%) were high risk, 39 (29.6%) were medium risk, and 9 (6.8%) were low risk. Of all the other inspected restaurants (the BASELINE set), 6225 (53.2%) were high risk, 2532 (21.5%) were medium risk, and 2967 (25.2%) were low risk (Table 1).

### Components of FINDER

FINDER estimates restaurant-level incidence rate of foodborne illness from web search and location data. It does so in a scalable and privacy-preserving way using two components: the web search model (WSM) identifies search queries about foodborne illness, and the location model (LM) identifies which restaurants have been visited by the relevant users. FINDER aggregates data at the restaurant level, and computes the proportion of users who visited each restaurant and later showed evidence of foodborne illness in their searches. We explain each step of this process in detail below.

### Web search model (WSM)

We developed a log-linear maximum entropy model that estimates, for an anonymized search query, the probability that the query is about foodborne illness. WSM training happens in a supervised way from automatically inferred labels. This allows us to deploy the model at scale and avoid relying on human raters, which can be very costly, and also allows us to maintain user privacy, as no live query is analyzed by humans.

In order to be able to automatically label training example queries, we focus on web pages about foodborne illness (broadly defined, including pertinent treatments and symptoms). We identify relevant web pages as those where concepts related to foodborne illness are prominently mentioned (this can be done using standard text classification techniques, which identify concept mentions in web pages).[27] Examples of such pages are the Wikipedia article about foodborne illness or the CDC web site devoted to foodborne illness. We observe that queries that lead to significant time spent on such pages are likely to truly be about foodborne illness, which allows us to label queries automatically. Anchoring on web pages allows us to regularize over the noise in individual queries, which—unlike pages—tend to be short, ambiguous, and often ungrammatical. The training pipeline automatically aggregates queries leading to relevant web sites, and uses them as positive examples. Then, it randomly samples other queries from the search stream to serve as negative examples. The WSM model is trained in a supervised way using these two (automatically labeled) sets of queries. The resulting model estimates the probability that a query is used for online research about foodborne illness (producing a score between 0 and 1 for each query), and does not require any human effort or inspection.

The model has a feature space of 50,000 dimensions, and leverages feature hashing for compactness. The features consist of word unigrams and bigrams extracted from the query string, as well as from the search result URLs, snippets (short summaries of each result displayed by the search engine), and web page titles. We also construct features based on Knowledge Graph[28] annotations of the concepts mentioned in the query.

Unlike much prior work,[12] FINDER estimates the actual incidence rate of foodborne illness in the population, rather than the overall query volume about it. That is, instead of computing the proportion of relevant search queries, FINDER computes the proportion of affected users. This distinction is important for two reasons. First, certain web users, such as medical professionals or academic researchers, may issue a significant number of pertinent queries, yet the plurality of their queries does not necessarily imply higher incidence of the disease. Second, focusing on users enables significantly better modeling of restaurant visits for users who opted in to use location history. In those cases, a user does not need to do anything specific at the restaurant to be included in the highly aggregated metrics. Prior work could only infer user's location if the user performed some online action (such as posting a Twitter message or doing a web search) at the venue or in the surrounding geographic area.[12,29] This requirement considerably limited the coverage of prior approaches, because only a minority of users actually take such a fortuitous action. FINDER does not have this constraint because it leverages ambient location that is collected in the background on mobile devices of opted-in users.

We applied the WSM query classifier to all English search queries in the United States to detect web searches related to foodborne illness, within an incubation period of 3 days after leaving a restaurant.

## Validation of the web search model

We validated the WSM's capacity to detect queries about foodborne illness via evaluation with human raters. To enable FINDER to learn at scale, no human labels were used for model training—we only collected human judgements for a relatively small sample of queries for evaluation. Since queries are inherently noisy, even experts may have different opinions. Therefore, we pooled judgements from multiple independent raters to obtain a more accurate estimate of ground truth. During evaluation, we compare labels predicted by the query classifier to this ground truth. The training and evaluation sets are by design disjoint and all queries across both sets are unique.

We sampled 15,000 queries for the evaluation and collected a total of 90,000 judgements on them (six independent judgements per query). The natural distribution of queries in the search stream clearly has a strong class imbalance for our task, whereas there are many fewer positive examples (queries related to foodborne illness) than negative ones. To address the challenge posed by this class imbalance and cover the full spectrum of positive as well as negative queries with a bounded human labeling budget, we up-sampled positive examples but otherwise mimicked the overall query distribution in order to accurately assess the performance of WSM on live data. To this end, half of the evaluation queries were sampled using a high-recall filter (designed to catch most foodborne illness queries), and the other half sampled using simple traffic weighting, where queries are sampled according to their frequency in the overall query stream. For the high-recall filter, we leveraged clicks on web pages about foodborne illness (annotated with Knowledge Graph topics). Specifically, we collected a large set of queries that led to clicks on such topical web pages, and then sampled queries out of this set according to their traffic weight. All queries were anonymized and highly aggregated to preserve privacy.

We employed two types of human raters: non-medical professionals as well as licensed medical doctors (MDs), trained in various medical specialties and located in the United States. Raters in both groups were unknown to and independent of the authors. Additionally, the raters were not aware of this research and did not know the purpose of the task. They were engaged by a third-party provider, also independent of the authors, which ensured proper qualifications of the raters.

Three non-medical professionals and three MDs independently judged the relevance of each search query in our sample to foodborne illness. The inter-rater agreement, computed over all judgements collected from both groups and measured by Krippendorff's alpha was 0.8, indicating high agreement. We aggregated all ratings from the six raters (three MDs and three non-professionals) for each query using the majority vote. Ties were broken using the majority rule over MD votes. We found this combination of raters produced the most accurate query labels, since MDs—experts in clinical diagnosis—are complemented by web raters who have a deeper experience with how health-related information needs could be reflected in search queries.

For each of the 15,000 queries in our evaluation set, we used WSM to predict the probability that the query is indicative of foodborne illness. This probability was then evaluated against the ground truth labels obtained from human raters as described above. In this evaluation, WSM achieved ROC AUC of 0.85 and F1 score of 0.74, which suggests it has high precision as well as high recall in identifying queries indicative of food poisoning.

## Location model (LM)

The location model (LM) connects the queries about foodborne illness, which were automatically extracted from web search logs, to restaurant visits extracted from location logs. The entire process is automated to preserve privacy, and the output signals are heavily aggregated. For each restaurant, FINDER estimates how many users visited it over the time period of interest. Next, FINDER uses WSM to compute the proportion of those visitors who searched for foodborne illness after the restaurant visit. This provides a probability estimate for a visitor to get infected within 3 days of visiting the restaurant. This period was selected based on the incubation periods of the most common foodborne illnesses,[9] as well as based on parameter optimization using historical inspection data. If a user visited more than one restaurant within that period, all visited restaurants were considered.

Thanks to aggregating data over numerous users, FINDER can confidently detect which restaurants are likely causing the illness, even though search and/or location evidence from any individual user may be ambiguous and noisy.

## Maintaining user privacy

The work reported herein has been conducted in accordance with the Google Privacy Policy and Terms of Service. At the beginning of processing, queries and locations have been anonymized using anonymous identifiers. This allowed FINDER to count the number of users who have visited a restaurant and later showed evidence of foodborne illness in their searches, in a privacy-preserving way using a differential privacy mechanism. All the processing has been done automatically, including the labeling of training examples for query classification (both positive and negative examples), so that no live query was analyzed by humans.

## Statistical analysis

To evaluate the ability of FINDER to detect unsafe restaurants as compared to BASELINE results, we used binomial logistic regression models with city and restaurant risk-level fixed effects to calculate odds ratios. We also used binomial logistic regression models to compare the performance of FINDER to COMPLAINT inspections and ROUTINE inspections. We used a linear regression model with city and restaurant risk-level fixed effects to calculate adjusted mean violation numbers. We used a multinomial logistic regression model to calculate relative risk ratios to compare the ability of FINDER to identify restaurants that received one of three grading results: Pass, Pass with Conditions, and Fail.

## Code availability

FINDER code was built on top of MapReduce open source code (https://github.com/GoogleCloudPlatform/appengine-mapreduce); however, the restaurant classification code cannot be published at this time.

## Data availability

The data that support the findings of this study were obtained from Google, Inc. and restrictions apply to the availability of these data, which were used under license for the current study, and so are not publicly available. Data may be available from authors upon reasonable request and with permission of Google, Inc.

## ACKNOWLEDGEMENTS

We thank Andrei Broder, Gerrin Butler, Susan Cadrecha, Stephanie Cason, Claire Cui, Jeff Dean, Jason Freidenfelds, John Giannandrea, Bryant Gipson, Benedict Gomes, Chad Heilig, Vivien Hoang, Anjali Joshi, Onur Kucuktunc, Scott Lee, Arthur Liang, Arne Mauser, Yael Mayer, Bhaskar Mehta, Leslie Miller, Caitlin Niedermeyer, Mike Pearson, Alvin Rajkomar, Karthik Raman, Thomas Roessler, John Sarisky, Calvin Seto, Tomer Shekel, Johanna Shelton, Diane Tang, Balder ten Cate, Chandu Thota, Kent Walker, Lawrence You, and health department inspectors for their help and advice. Data for this paper were obtained from Google databases as well as publicly available data sets from the Southern Nevada Health District (https://southernnevadahealthdistrict.org/restaurants/inspections.php) and the Chicago Department of Public Health (https://data.cityofchicago.org/Health-Human-Services/Food-Inspections-Dashboard/2bnm-jnvb). This publication was supported by the Cooperative Agreement Number 1U01EH001301-01 funded by the Centers for Disease Control and Prevention. Its contents are solely the responsibility of the authors and do not necessarily represent the official views of the Centers for Disease Control and Prevention or the Department of Health and Human Services.

## AUTHOR CONTRIBUTIONS

A.S., P.R., and E.G. conceived and implemented FINDER. A.S., L.D., M.B., R.M., T.S., P.R., and E.G. designed deployment of FINDER. L.D. and M.B. oversaw implementation of FINDER in Las Vegas. R.M and T.S. oversaw implementation of FINDER in Chicago. S.C. and A.K.J. analyzed data and performed statistical analysis. All authors contributed to writing the manuscript. A.S. and S.C. contributed equally to this manuscript and therefore are listed as co-first authors.

## ADDITIONAL INFORMATION

**Supplementary information** accompanies the paper on the *npj Digital Medicine* website (https://doi.org/10.1038/s41746-018-0045-1).

**Competing interests:** A.S., P.R., and E.G. are directly employed by Google. The remaining authors declare no competing interests.

**Publisher's note:** Springer Nature remains neutral with regard to jurisdictional claims in published maps and institutional affiliations.